\documentclass[lettersize,journal]{IEEEtran}
\usepackage{amsmath,amsfonts, amssymb}
\usepackage{algorithmic}
\usepackage{algorithm}
\usepackage{array}
\usepackage{bbm}
\usepackage[caption=false,font=footnotesize,labelfont=rm,textfont=rm,subrefformat=parens]{subfig}
\usepackage{textcomp}
\usepackage{stfloats}
\usepackage{url}
\usepackage{verbatim}
\usepackage{graphicx}
\usepackage{cite}
\usepackage{multirow}
\usepackage{setspace}
\usepackage{diagbox}
\usepackage{booktabs}
\usepackage{makecell}

\newcommand{\Statex}{\item[]}

\usepackage[colorlinks  = true,
            linkcolor   = black,
            urlcolor    = blue,
            anchorcolor = blue,
            bookmarks   = false,
            citecolor   = blue,
            ]{hyperref}

\begin{document}

\title{CSIYOLO: An Intelligent CSI-based Scatter Sensing Framework for Integrated Sensing and Communication Systems}

\author{Xudong Zhang,~\IEEEmembership{Graduate Student Member,~IEEE}, Jingbo Tan,~\IEEEmembership{Member,~IEEE}, Zhizhen Ren,\\  Jintao Wang,~\IEEEmembership{Senior Member,~IEEE}, Yihua Ma,~\IEEEmembership{Member,~IEEE}, and Jian Song,~\IEEEmembership{Fellow,~IEEE}
\thanks{  
This work was supported in part by the Beijing Natural Science Foundation (Grant No. L242083) and in part by the National Natural Science Foundation of China (Grant No. 62401315). \textit{(Corresponding author: Jingbo Tan, Jintao Wang.)}

The open-source codes of this paper are available at \href{https://github.com/zhang-xd18/csiyolo}{https://github.com/zhang-xd18/csiyolo}.

Xudong Zhang, Jingbo Tan, Zhizhen Ren, and Jintao Wang are with the Department of Electronic Engineering, Tsinghua University, Beijing 100084, China, and Beijing National Research Center for Information Science and Technology (BNRist) (e-mail: zxd22@mails.tsinghua.edu.cn; tanjingbo@tsinghua.edu.cn; rzz21@mails.tsinghua.edu.cn; wangjintao@tsinghua.edu.cn).

Yihua Ma is with the State Key Laboratory of Mobile Network and Mobile Multimedia Technology, Shenzhen, China, and the ZTE Corporation Beijing Research Institute, Beijing, China (e-mail: yihua.ma@zte.com.cn).

Jian Song is with the Shenzhen International Graduate School, Tsinghua University, Shenzhen 518055, China (e-mail: jsong@tsinghua.edu.cn).
}
}

\maketitle
\begin{abstract}
Integrated sensing and communication (ISAC) is regarded as a promising technology for next-generation communication systems, enabling simultaneous high-rate data transmission and accurate target sensing. 
Among various tasks in ISAC, scatter sensing plays a crucial role in exploiting the full potential of ISAC and supporting applications such as autonomous driving and low-altitude economy.
However, most existing methods rely on either waveform and hardware modifications or traditional signal processing schemes, leading to poor compatibility with current communication systems and limited sensing accuracy.
To address these challenges, we propose CSIYOLO, a framework that performs scatter localization only using estimated channel state information (CSI) from a single base station-user equipment pair.
This framework comprises two main components: anchor-based scatter parameter detection and CSI-based scatter localization.
First, by formulating scatter parameter extraction as an image detection problem, we propose an anchor-based scatter parameter detection method inspired by You Only Look Once architectures, with an adaptive ghost target removal strategy.
After that, a CSI-based localization algorithm is derived to determine scatter locations with extracted parameters.
Moreover, to improve localization accuracy and implementation efficiency, we design an extendable network structure with task-oriented optimizations, enabling multi-scale anchor detection and better adaptation to CSI characteristics. 
A noise injection training strategy is further designed to enhance robustness against channel estimation errors.
Since the proposed framework operates solely on estimated CSI without modifying waveforms or signal processing pipelines, it can be seamlessly integrated into existing communication systems as a plugin.
Experiments show that our proposed method can significantly outperform existing methods in scatter localization accuracy with relatively low complexities under varying numbers of scatters and estimation errors.

\end{abstract}

\begin{IEEEkeywords}
Environment sensing, scatter localization, deep learning, integrated sensing and communications
\end{IEEEkeywords}

\section{Introduction}
\subsection{Background and motivation}
Integrated sensing and communication (ISAC) has been widely recognized as a key enabler for next-generation communication systems, attracting considerable research attention in recent years.
As one of the six potential technologies of 6G identified by the International Mobile Telecommunication (IMT) 2030 framework \cite{Wei2023Survey}, ISAC aims to simultaneously support effective data transmission and accurate object sensing. 
With increasing carrier frequencies and advanced signal designs, the potential of using communication signals for sensing has been extensively explored, opening new opportunities for ISAC development \cite{fan2024isac, fan2024tvt}.
Significant efforts have been devoted to extracting sensing information from ISAC signals, including designing waveforms with favorable correlation properties and embedding additional sensing signals.
Building on these advances, ISAC is paving the way for emerging applications such as the low-altitude economy \cite{jiang2025low}, autonomous driving \cite{roos2019radar}, and the intelligent Internet of Things \cite{cui2021iot}.

Among various tasks in ISAC, scatter sensing is particularly important to further unlock the full potential of ISAC.
On the one hand, scatter sensing, encompassing three-dimensional (3D) environment reconstruction and scatter localization, can provide comprehensive environmental awareness, which facilitates more intelligent connectivity and efficient transmission in practical applications such as the low-altitude economy \cite{jiang2025low}.
On the other hand, scatter sensing can bring great benefits to the optimization of communication systems through operations such as channel management, resource allocation, and adaptive signal processing.
Environment-assisted communication optimization has been motivated by ISAC and attracted extensive research recently, including environment-assisted beam prediction \cite{environment2025feng}, blockage control \cite{charan2022computer}, and channel reconstruction \cite{zhang2024icc}.
These studies demonstrate the potential of leveraging environmental information to enhance communication performance, laying the foundation for more intelligent and adaptive communication systems.

However, the limited sensing capability and information loss inherent in communication-oriented signal designs of current communication systems still pose great obstacles to sensing performance, especially for scatter sensing.
Consequently, how to achieve effective scatter sensing with existing communication signals has become a critical issue in ISAC research.

\subsection{Related works}
Scatter sensing has attracted considerable attention in ISAC research.
Conventional approaches primarily rely on radar systems or dedicated waveforms, which are often costly and challenging to integrate into existing communication infrastructures. 
For example, frequency-modulated continuous wave (FMCW) waveforms are widely adopted in environmental reconstruction research \cite{santra2020ambiguity}, where echo signals are typically processed using discrete Fourier transform (DFT)-based algorithms \cite{wei2023iterative} or multiple signal classification (MUSIC) methods \cite{kim2019MUSIC, chen2023music,lu2024isac4d}. 
However, the resolution limitations of such techniques and the inherent constraints of FMCW waveforms have been identified in some cases \cite{niu2025interference, meng2022adaptive}.
In addition, radar waveforms generally entail high hardware costs and integration challenges, leading to poor compatibility with widespread communication systems. 
Consequently, leveraging existing communication signals—such as orthogonal frequency-division multiplexing (OFDM)—for scatter sensing has emerged as a promising direction, offering greater flexibility and compatibility for ISAC deployment.

In recent years, substantial progress has been made in scatter sensing using existing communication signals, mainly through two approaches: transmission delay-based methods across multiple communication nodes, and multi-dimensional signal processing schemes in massive multiple-input multiple-output (MIMO)-OFDM systems.

On one hand, the transmission delay-based methods estimate scatter positions from the difference of signal transmission delays among multiple nodes, such as base stations (BSs) and user equipments (UEs). 
This has motivated a body of work on scatter localization using OFDM signal delay analysis across multiple nodes \cite{guo2024user, shi2022devicefree, shi2023joint, wang2022trilateration}. 
For instance, \cite{shi2022devicefree} proposes a device-free localization technique based on time-of-flight (ToF) estimation, employing a two-phase framework to determine estimation ranges and remove ghost targets. 
Such methods typically require collaboration among multiple nodes to reduce ambiguity \cite{guo2024user, shi2023joint}.
However, effectively associating scatter clusters and eliminating ghost targets remains challenging. 
Furthermore, the number of available collaboration nodes and estimation errors of individual nodes can significantly degrade localization accuracy.

On the other hand, with the widespread deployment of massive MIMO-OFDM systems, scatter localization can be achieved with a single BS-UE pair by exploiting spatial information like angle-of-arrivals.
Advanced signal processing algorithms have been developed to extract environmental information from multi-dimensional ISAC signals in such systems.
For example, researchers in \cite{xu2023jointlong, huang2022joint} propose advanced optimization algorithms to perform joint scatter localization and channel estimation based on downlink and uplink pilot signals.
Similarly, \cite{Baquero2022mmwave} employs millimeter-wave signals for scatter localization through continuous tracking in the detected area.
Moreover, the electromagnetic properties of scatters are investigated from received OFDM echoes in \cite{jiang2024electro}.
Nevertheless, most of these approaches require specific signal modifications or additional hardware adjustments to the transceivers, leading to increased implementation costs.

To avoid modifications to signal structures or hardware, recent studies have explored using estimated CSI for scatter sensing, offering a more flexible and compatible solution with a single BS-UE pair.
This approach leverages the fact that environmental propagation characteristics are inherently embedded in CSI, allowing them to be extracted seamlessly without altering existing communication pipelines \cite{kadambi2022icc}.
For example, \cite{mou2023mmwave} introduces a low-rank tensor decomposition (LTD) method to extract scatter parameters including path transmission delays and angles from CSI matrices, and realize path classification using neural networks. 
Similarly, \cite{lu2024isac4d} leverages both uplink and downlink CSI for four-dimensional environmental reconstruction by first estimating scatter parameters with MUSIC and DFT-based methods, and then fusing extracted features via a dedicated network to enhance reconstruction accuracy. 
In addition, \cite{song2024environment} investigates the influence of channel properties, such as frequency and bandwidth, on CSI-based sensing performance.
However, the aforementioned approaches remain dependent on conventional parameter estimation algorithms like MUSIC and DFT, whose performance deteriorates severely in scenarios with varying numbers of scatters or under noisy channel conditions \cite{tran2018bayesian}.

In summary, current scatter sensing approaches face two main limitations: (i) poor compatibility with existing communication pipelines due to multi-node collaboration or modifications to hardware, waveforms, and signal processing, and (ii) degraded accuracy caused by resolution limits, ghost targets, and sensitivity to noisy channels or variable scatter numbers.
Therefore, to cope with these limitations, it is essential to develop robust and adaptable scatter localization methods that can be seamlessly integrated into existing communication systems, which remains an important and unresolved issue.

\vspace{-3mm}
\subsection{Contributions}
To address these challenges, this paper proposes a novel scatter sensing framework leveraging estimated CSI in MIMO-OFDM systems with a single BS-UE pair. 
This framework formulates scatter localization as an object detection problem and develops an efficient anchor-based detection method inspired by You Only Look Once (YOLO) architectures \cite{Redmon2016yolo}, with an adaptive ghost target removal strategy. 
Network design and training strategies are optimized to exploit the feature distribution characteristics of CSI, improving robustness against varying scatter numbers and noisy channel conditions.
The contributions of this paper are summarized as follows:
\begin{itemize}
    \item We propose a pluggable CSI-based scatter sensing framework called CSIYOLO that utilizes estimated CSI for scatter localization and can be seamlessly integrated into existing communication signal processing pipelines. The framework consists of two stages: anchor-based scatter parameter detection and CSI-based scatter localization. First, an anchor-based detection method is designed to estimate scatter parameters from CSI matrices with adaptive ghost target removal, and then a localization algorithm with a single BS-UE pair is proposed to achieve precise scatter positioning based on extracted parameters.
    \item Scatter parameter estimation is formulated as an object detection task, for which we design an anchor-based detection network inspired by YOLO. 
    Instead of directly applying YOLO to image detection, we leverage the unique characteristics of CSI and redesign the detection logic and loss functions to meet the requirements of scatter sensing.
    The network captures the continuous mapping between CSI distribution and scatter parameters, while adaptively estimating detection confidence to remove ghost targets, achieving accurate localization under variable scatter numbers.
    \item To improve localization accuracy and robustness under noisy channels, an extendable network structure with task-oriented optimizations is carefully designed with multi-scale anchor detection. We introduce the circular convolution for better feature extraction in the angular domain, and utilize convolution factorization to enable scalable and efficient implementation. Additionally, a noise injection training strategy is incorporated to enhance robustness against channel estimation errors.
    \item Extensive experiments demonstrate the effectiveness of CSIYOLO over conventional methods. The framework achieves higher localization accuracy with relatively low complexities under variable scatter numbers, successfully removes ghost targets, and maintains performance under noisy channels, significantly outperforming traditional approaches that even assume prior knowledge of the exact number of scatters.
\end{itemize}

\vspace{-2mm}
\subsection{Organization and notations}
The remainder of this paper is organized as follows.
Section II introduces the ISAC system model and formulates the scatter sensing task.
Section III presents the proposed intelligent CSI-based scatter sensing framework, including the anchor-based scatter parameter detection and CSI-based scatter localization.
Section IV details the efficient and extendable network design along with the noise injection training strategy.
In Section V, we present the simulation settings and experimental results to validate the effectiveness of the proposed method.
Finally, Section VI concludes the paper.

\textbf{Notations}:
Bold uppercase letters, bold lowercase letters, and normal font represent matrices, vectors, and scalars, respectively.
$(\cdot)^{\text{T}}$ and $(\cdot)^{\text{H}}$ denote transpose and Hermitian transpose, respectively.$\mathbf{A}[i,j]$ is the $(i,j)$-th element of matrix $\mathbf{A}$. $|a|$ and $\Vert \mathbf{a} \Vert$ denote the absolute value of a scalar $a$ and the $\ell_2$ norm of vector $\mathbf{a}$, respectively.
$\lfloor \cdot \rfloor$ denotes the floor function, $[\cdot]\times[\cdot]$ the Cartesian product of intervals, and $\mathbbm{1}$ the indicator function.
$\{\cdot\}$ represents a set.
The cardinality of a set $\mathbb{A}$ is denoted by $|\mathbb{A}|$.
$\mathbb{C}$, $\mathbb{R}$, and $\mathbb{Z}$ denote the sets of complex, real, and integer numbers, respectively.
$\mathcal{N}(\mu,\sigma^2)$ represents a Gaussian distribution with mean $\mu$ and variance $\sigma^2$.

\section{System model}
In this section, we present the signal model of the ISAC system and formulate the CSI-based scatter sensing problem for a MIMO-OFDM system with a single BS-UE pair.

\subsection{ISAC system}

\begin{figure}[!t]
\centering
\includegraphics[width=0.9\linewidth]{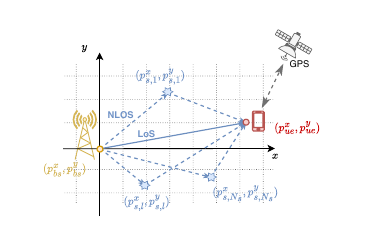}
\vspace{-2mm}
\caption{The diagram of the ISAC system. $N_s$ scatters are randomly distributed in the communication channel between the BS and the UE, with both LoS and NLoS paths. Besides, the UE position can be obtained through GPS.}
\vspace{-3mm}
\label{fig_isac_system} 
\end{figure}

As shown in Fig. \ref{fig_isac_system}, we consider a single-cell ISAC system comprising a single BS-UE pair. 
For multi-user scenarios, the proposed method can be readily applied to each user individually, with future work focusing on effective merging strategies to aggregate multi-user information for improved sensing accuracy.
The BS is equipped with $N_t$ transmit antennas under the uniform linear array (ULA) configuration, and the UE is equipped with $N_r$ receive antennas.
For simplicity, $N_r$ is set to 1.
The OFDM waveform is utilized for data transmission with $N_c$ sub-carriers.
A total of $N_s$ scatters are randomly distributed in the wireless propagation environment between the BS and UE, forming the downlink communication channel and serving as the sensing targets.
During pilot training, the UE estimates the downlink channel to facilitate data reception and demodulation. These estimated channels inherently contain information about the environmental scatters, thereby enabling the possibility of performing environment sensing.

Specifically, consider the BS is located at a known position $\mathbf{p}_{bs} = \left[p_{bs}^x, p_{bs}^y\right]$ and the UE is located at $\mathbf{p}_{ue} = \left[p_{ue}^x, p_{ue}^y\right]$, which can be obtained with the Global Positioning System (GPS) in the UE \cite{guo2024user}.
Besides, for the $l$-th scatter, its position is denoted as $\mathbf{p}_{s,l} = [p_{s,l}^x, p_{s,l}^y]$. \footnote{For more practical 3D scenarios, the uniform planar array is required in the BS to obtain the elevation angle of departure. In this case, the following system formulations can be readily extended with the $z$-coordinate. Likewise, our proposed method can be adapted with 3D derivations and convolutions. Detailed formulations for 3D scenarios are left for future investigation.}
Thus, the downlink communication channel on the $n_c$-th sub-carrier can be formulated based on the Saleh-Valenzuela model \cite{saleh1987statistical} as:
\begin{equation}
    \mathbf{h}_{n_c} = \sum_{l=0}^{N_s} \alpha_l e^{-j 2 \pi \tau_l (f_0 + n_c \Delta f)} \mathbf{a}_T\left(\theta_l\right), 
\label{channel}
\end{equation}
where $\alpha_l$, $\tau_l$, and $\theta_l$ represent the complex channel gain, transmission delay, and angle of departure (AoD) of the $l$-th path, respectively.
$f_0$ and $\Delta f$ denote the center frequency and sub-carrier spacing, respectively.
$\mathbf{a}_T\left(\theta\right)$ is the steering vector of the transmit antennas at the BS, which can be expressed as
\begin{equation}
\label{steer_vector}
\mathbf{a}_T(\theta_l) = \left[1, e^{-j 2\pi \frac{d \sin(\theta_l)}{\lambda}}, ..., e^{-j 2\pi \frac{(N_t - 1) d \sin(\theta_l)}{\lambda}}\right]^\text{T},
\end{equation}
where $d$ is the antenna spacing and $\lambda$ is the wavelength of the communication signal.
Note that $l = 0$ represents the line-of-sight (LoS) communication path that typically has a larger channel gain than the other $N_s$ non-LoS (NLoS) paths.

Through denoting $\mathbf{p}_{ue}$ as $\mathbf{p}_{s,0}$, the delay $\tau_l$ and the AoD $\theta_l$ of the $l$-th path ($0 \leq l \leq N_s$) can be further expressed as:
\begin{equation} 
\label{delay}
\begin{split}
\tau_l = \frac{\Vert \mathbf{p}_{s,l} - \mathbf{p}_{bs} \Vert + \Vert \mathbf{p}_{s,l} - \mathbf{p}_{ue} \Vert - \Vert \mathbf{p}_{bs} - \mathbf{p}_{ue} \Vert}{c} &, \\
\theta_l = \arctan\left(\frac{p_{s,l}^y - p_{bs}^y}{p_{s,l}^x - p_{bs}^x}\right) ~~~~~~~~~&,
\end{split}
\end{equation}
where $c$ denotes the speed of light. 
Finally, the channel matrix of all the sub-carriers can be expressed as $\mathbf{H} = \left[\mathbf{h}_0^{\text{H}}; \mathbf{h}_1^{\text{H}}; ...; \mathbf{h}_{N_c - 1}^{\text{H}}\right] \in \mathbb{C}^{N_c \times N_t}$.
Correspondingly, the received signal on the $n_c$-th sub-carrier in the UE can be modeled as:
\begin{equation}
\label{receive_signal}
y_{n_c} = \mathbf{h}_{n_c}^{\text{H}} \mathbf{x}_{n_c} + z_{n_c},
\end{equation}
where $\mathbf{x}_{n_c} \in \mathbb{C}^{N_T \times 1}$ is the transmitted symbols carried on the $n_c$-th sub-carrier and $z_{n_c}$ is the additive Gaussian noise.
In traditional communication signal processing pipelines, the BS typically performs pilot training to enable channel estimation at the UE.
With conventional channel estimation methods like least square or minimum mean square error estimation, the UE can obtain the estimated channel matrix:
\begin{equation}
\label{channel_estimation}
\hat{\mathbf{H}} = \text{CE}\left(\mathbf{y}, \mathbf{X^p}\right),
\end{equation}
where $\mathbf{y} = \left[y_0, y_1, ..., y_{N_c - 1}\right]$ is the received signal vector and $\mathbf{X^p} = \left[\mathbf{x}^p_0, \mathbf{x}^p_1, ..., \mathbf{x}^p_{N_c - 1}\right]$ is the transmitted pilot symbols. 
For the sake of simplicity, we will uniformly use $\mathbf{H}$ to represent the estimated channel matrix in the following parts.

\subsection{CSI-based scatter sensing}
As shown in (\ref{delay}), the scatter positions determine the CSI through scatter parameters like transmission delays and AoDs, enabling the possibility of inferring scatter positions from estimated CSI.
Thus, the CSI-based scatter sensing problem with a single BS-UE pair can be formulated as \cite{guo2024user}:
\begin{equation}
\label{scatter_sensing}
\left\{\mathbf{p}_{s,l} \right\}_{1 \leq l \leq N_s} = f\left(\mathbf{H}, \mathbf{p}_{bs}, \mathbf{p}_{ue}\right),
\end{equation}
where the mapping function $f(\cdot)$ is the CSI-based scatter sensing method to be designed.

To improve the scatter localization accuracy, there are several challenges to be addressed.
\begin{itemize}
    \item \textbf{Variable number of scatters}. Traditional methods like MUSIC always perform well in case the number of scatters is known in advance. However, this assumption is often invalid in practical scenarios where the number of scatters is unknown and may vary.
    \item \textbf{Channel estimation errors}. The CSI estimated in the UE inevitably suffers from information loss caused by estimation errors, making information extraction challenging and reducing the accuracy of scatter localization.
    \item \textbf{Ghost target removal}. Due to the influence of multi-path effects, some ghost targets may be mis-detected as true scatters and greatly degrade the localization performance.
\end{itemize}

However, most existing CSI-based methods rely on traditional parameter estimation techniques such as MUSIC and DFT, which are sensitive to channel estimation errors and ghost target interference, and often struggle to handle scenarios with variable numbers of scatters \cite{tran2018bayesian}.

\vspace{-3mm}
\section{CSIYOLO: an intelligent CSI-based scatter sensing framework}
\begin{figure*}[!t]
    \centering
    \includegraphics[width=0.8\linewidth]{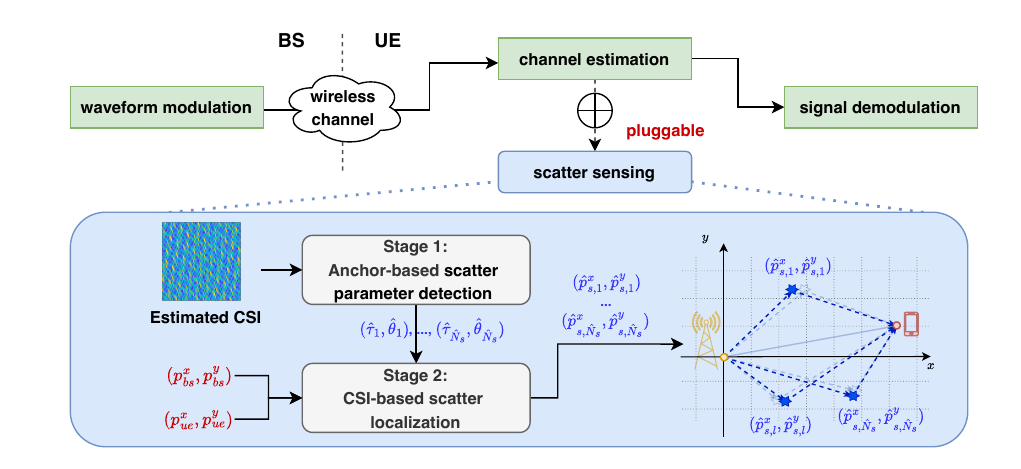}    
    \vspace{-2mm}
    \caption{The main structure of the CSIYOLO framework. It consists of two main stages. First, the estimated CSI is processed by the anchor-based scatter parameter detection method to extract scatter parameters. Secondly, the scatter parameters are used to localize the scatters based on the ellipse model.}
    \label{Fig_framework}
    \vspace{-2mm}
\end{figure*}

To address these challenges, we propose an intelligent CSI-based scatter sensing framework named CSIYOLO, which can effectively localize scatters with unknown quantities and enhanced robustness against channel estimation errors.
As illustrated in Fig. \ref{Fig_framework}, the proposed framework can be seamlessly integrated into existing communication systems as a pluggable module. It comprises two main stages: anchor-based scatter parameter detection and CSI-based scatter localization.

\vspace{-1mm}
\subsection{Stage 1: Anchor-based scatter parameter detection}

In this subsection, we detail the proposed anchor-based scatter parameter detection method. 
First, the scatter parameter extraction task is formulated as an object detection problem. 
Subsequently, an anchor-based detection method is designed to accurately extract scatter parameters from the channel while adaptively removing ghost targets.

\subsubsection{Parameter extraction as an object detection task}
Instead of traditional methods that directly perform parameter estimation on the channel matrix through likelihood or grid search approaches, we transform the parameter extraction problem into an object detection task. 
First, the channel matrix in the frequency-antenna domain $\mathbf{H}$ is transferred into the angular-delay domain $\bar{\mathbf{H}}$ with two-dimensional (2D) inverse DFT as:
\begin{equation}
\label{DFT}
\bar{\mathbf{H}} = \mathbf{F}_a \mathbf{H} \mathbf{F}_d,
\end{equation}
where $\mathbf{F}_a \in \mathbb{C}^{N_c \times N_c}$ and $\mathbf{F}_d \in \mathbb{C}^{N_t \times N_t}$ are DFT matrices.
Through transformation, the channel sparsity is improved and distinct transmission paths are distributed separately across the 2D plane, with unavoidable overlap and interference.

\begin{figure}[!t]
\includegraphics[width=\linewidth]{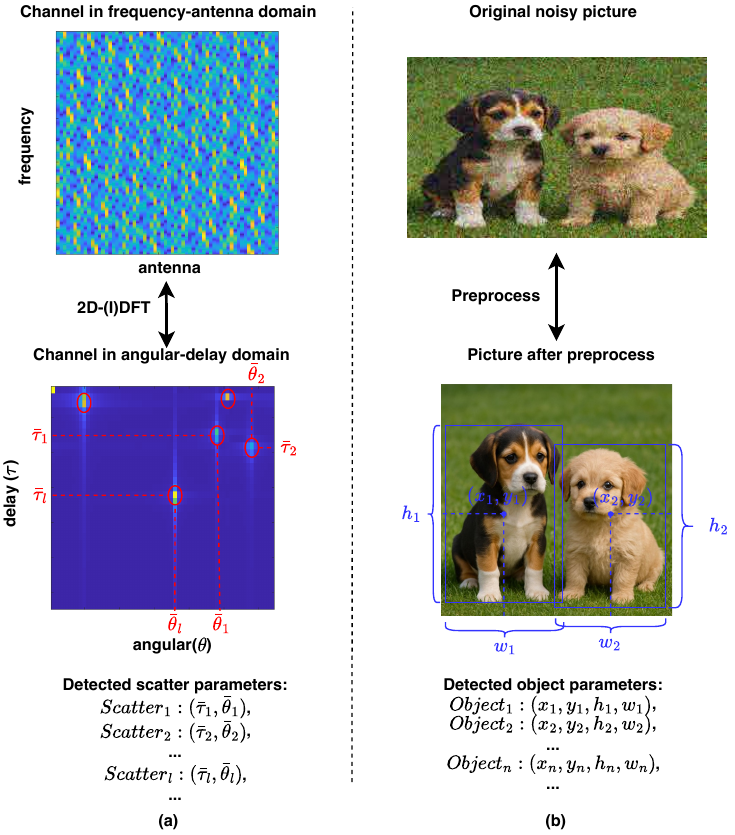}
\vspace{-7mm}
\caption{The diagram of the channel parameter extraction as an object detection task. (a) Scatter parameter extraction. The original channel in the frequency-antenna domain is transformed into the angular-delay domain, allowing scatter parameters to be extracted from the coordinates $(\bar{\tau}, \bar{\theta})$ of the objects. (b) Object detection in the CV field. The original noisy image is first pre-processed to resize and remove noise. Subsequently, the objects in the image are detected with their positions $(x,y)$ and shapes $(h,w)$.}
\vspace{-5mm}
\label{fig_channel} 
\end{figure}

As shown in Fig. \ref{fig_channel}(a), the scatters, i.e., transmission paths, can be identified by their 2D coordinates on the angular-delay plane, corresponding to their transmission AoDs and delays.
In this case, the estimation of the scatter parameters $\theta_l$ and $\tau_l$ can be transformed into an object detection and localization problem over the 2D plane, trying to get the 2D scatter coordinates.
Specifically, considering the discrete channel matrix in the frequency-antenna domain, $\mathbf{H}$ can be expressed as:
\vspace{-1mm}
\begin{equation}
    \label{discrete_channel}
    \mathbf{H}\left[m ,n\right] = \sum_{l=0}^{N_s} \alpha_l e^{-j 2 \pi \tau_l (f_0 + m \Delta f)} e^{- j 2\pi n \frac{d \sin(\theta_l)}{\lambda}}.
\end{equation}

After the 2D inverse DFT, the channel matrix in the angular-delay domain can be expressed as:
\vspace{-1mm}
\begin{equation}
    \label{after_DFT_eq}
    \begin{split}
        \bar{\mathbf{H}}\Big[\bar{\tau}, \bar{\theta}\Big] & = \sum_{m=0}^{N_c - 1} \sum_{n=0}^{N_t - 1} \mathbf{H}\left[m, n\right] e^{j 2\pi \frac{m \bar{\tau}}{N_c}} e^{j 2\pi \frac{n \bar{\theta}}{N_t}} \\
            = \sum_{l=0}^{N_s} & \alpha_l^{'} \sum_{m=0}^{N_c - 1} e^{-j 2\pi m (\Delta f \tau_l - \frac{\bar{\tau}}{N_c})} \sum_{n=0}^{N_t - 1} e^{-j 2\pi n (\frac{d \sin(\theta_l)}{\lambda} - \frac{\bar{\theta}}{N_t})} \\
            = \sum_{l=0}^{N_s} & \alpha_l^{'}  g_1(\Delta f \tau_l, \bar{\tau}) g_2\left(\frac{d \sin(\theta_l)}{\lambda}, \bar{\theta}\right),
    \end{split}
\end{equation}
where $\alpha_l^{'} = \alpha_l e^{-j 2\pi f_0 \tau_l}$, and $\left(\bar{\tau}, \bar{\theta}\right)$ is the scatter coordinate in the 2D angular-delay domain.

Considering the widely-used settings $d = \frac{\lambda}{2}$, the coordinate of the $l$-th scatter ``object'' in the angular-delay domain can be derived through maximizing the modulus of $\bar{\mathbf{H}}$:
\begin{equation}
    \label{after_DFT}
    \begin{split}
        \bar{\tau}_l & = \left\lfloor \Delta f \tau_l N_c \right\rfloor, \\
        \bar{\theta}_l & = \left\lfloor \left(2 Z + \frac{\sin(\theta_l)}{2}\right) N_t \right\rfloor, Z \in \mathbb{Z} \\
        & = \begin{cases}
            & \left\lfloor \frac{\sin(\theta_l)}{2} N_t \right\rfloor, \sin(\theta_l) \geq 0, \\[5pt]
            & \left\lfloor\left(\frac{\sin(\theta_l)}{2} + 1\right) N_t \right\rfloor, \sin(\theta_l) < 0.
        \end{cases}
    \end{split}
\end{equation}

With (\ref{after_DFT}), the scatter parameters $\left(\tau_l, \theta_l\right)$ can be further derived from the coordinates $\left(\bar{\tau}_l, \bar{\theta}_l\right)$ as:
\begin{equation}
    \label{parameter_from_coordinate}
    \begin{split}
        \tau_l & = \frac{\bar{\tau}_l}{\Delta f N_c}, \\
        \theta_l & = \arcsin\left[\frac{2 \bar{\theta}_l}{N_t} - 2 \cdot \mathbbm{1}\left(\bar{\theta}_l > \frac{N_t}{2}\right)\right].
    \end{split}
\end{equation}
It should be noted that although the coordinates on the 2D grid are integers, our proposed method can infer continuous coordinate positions, overcoming the precision limitations imposed by the grid. 
By interpreting the angular-delay domain channel matrix as a 2D image and scatter transmission paths as objects within it, the scatter parameter extraction task can be naturally transformed into an object detection problem, as illustrated in Fig. \ref{fig_channel}, which is similar to the well-studied object detection problem in computer vision (CV).

\subsubsection{YOLO-inspired scatter detection with anchors}
With the rapid development of CV, object detection techniques have gone through the evolution from two-stage detection methods like Fast R-CNN \cite{Girshick2015fast} to the one-stage detection method like YOLO series \cite{Redmon2016yolo, redmon2018yolov3, yolov5github}.
Practices have shown the superiority of the YOLO series in terms of both detection accuracy and efficiency over traditional two-stage methods, achieving real-time detection with high accuracy.

However, the detection of scatter coordinates in the angular-delay domain CSI matrix still has great differences compared to the traditional object detection task in CV:
\begin{itemize}
    \item \textbf{Object shape.} 
    In the angular-delay domain CSI matrix, the objects differ from the rectangular bounding boxes commonly assumed in most CV object detection tasks. Instead, they exhibit a spreading pattern along both dimensions, originating from a central point. Although the spread may be relatively large, the path energy is primarily concentrated near the center. Consequently, the conventional CV object detection logic of capturing object boundaries is not directly applicable in this context.
    \item \textbf{Detection purpose.} 
    The objective of scatter parameter detection is to accurately identify the 2D coordinates of the scatter object centers, which contrasts with the traditional object detection goal of determining object bounding boxes. In the CSI matrix, bounding boxes are neither well-defined nor relevant to scatter localization. Therefore, directly applying traditional object detection methods may lead to unnecessary computational overhead and suboptimal performance.
\end{itemize}

Thus, task-oriented optimizations are necessary for scatter parameter detection beyond conventional object detection methods.
Accordingly, we propose an anchor-based detection method inspired by the anchor and regression concepts of YOLO, which can greatly improve the detection accuracy and efficiency.

\begin{figure}[!t]
\centering
\includegraphics[width=0.85\linewidth]{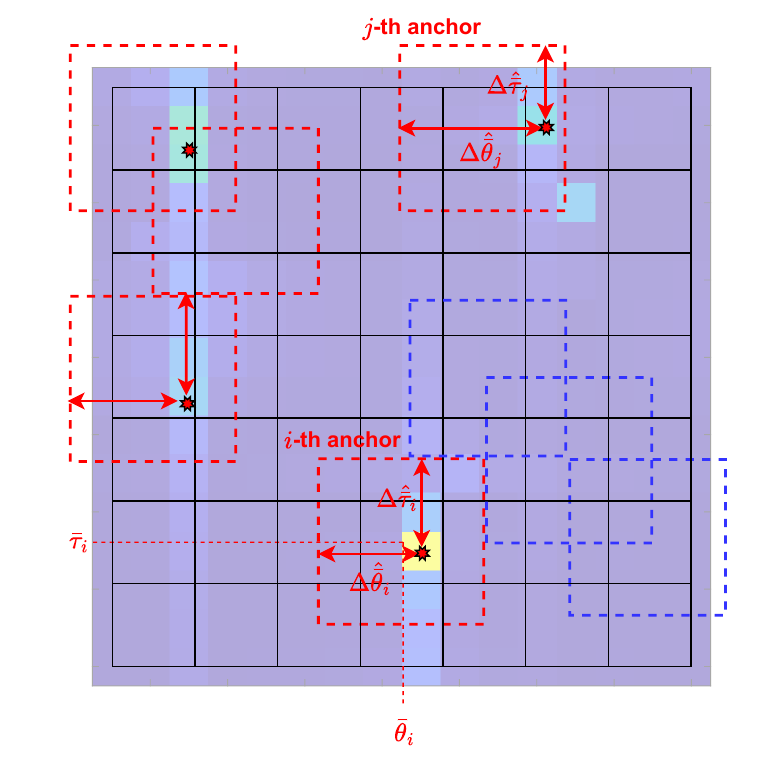}
\vspace{-2mm}
\caption{The diagram of the anchor-based scatter parameter detection method. Dashed boxes represent the scanning regions of the anchors. The red box and blue box represent the positive anchor and negative anchor, respectively. }
\vspace{-3mm}
\label{fig_anchor} 
\end{figure}

As shown in Fig. \ref{fig_anchor}, instead of directly projecting the channel matrix to the estimated parameters, we formulate each element in the channel matrix as an ``anchor'' in the proposed detection method.
This means that each element will perform as an estimator for the nearby scatter coordinates.
It will estimate the relative positions of the nearby scatters with the confidence of this estimate.
Through merging outputs of all the anchors, all the scatter coordinates can be obtained.
Specifically, for the $i$-th anchor of the channel matrix, corresponding to the matrix element at the coordinate $\left(\bar{\tau}_i, \bar{\theta}_i\right)$, it will scan a nearby region in the matrix like $\left[\bar{\tau}_i - \Delta \bar{\tau} / 2, \bar{\tau}_i + \Delta \bar{\tau} / 2\right] \times \left[\bar{\theta}_i - \Delta \bar{\theta} / 2, \bar{\theta}_i + \Delta \bar{\theta} / 2\right]$.
Within this region, the anchor will estimate the relative offset $\left(\Delta \hat{\bar{\tau}}_i, \Delta \hat{\bar{\theta}}_i\right)$ of the nearby scatter with respect to a reference point (typically the upper-left corner).
In this way, the coordinates of the scatters can be estimated continuously, overcoming accuracy limitations caused by grid resolution.
With relative coordinate offsets, the absolute scatter coordinates can be easily derived by adding them to the reference point coordinates.
Meanwhile, the anchor also estimates and outputs the confidence level $\hat{C}_i$ of this detected scatter.
Based on the detection method, we design an anchor-based network $f(\cdot)$ to process the channel $\bar{\mathbf{H}}$ and output the detection results of all the anchors as:
\begin{equation}
\label{forward}
\left\{\left(\Delta \hat{\bar{\tau}}_i, \Delta \hat{\bar{\theta}}_i, \hat{C}_i\right)\right\}_{1 \leq i \leq N_{a}} = f(\bar{\mathbf{H}}),
\end{equation}
where $N_a$ is the number of anchors determined by the output size of the network, which will be detailed in Section IV.

Thus, the training target of a single anchor at $\left(\bar{\tau}_i, \bar{\theta}_i\right)$ with the ground truth of the scatter coordinate offset $\left(\Delta \bar{\tau}_i, \Delta \bar{\theta}_i\right)$ consists of two parts: the localization loss $L_{loc}$ and the objectness loss $L_{obj}$, as follows,
\begin{equation}
    \label{training_target}
    \begin{split}
        L(\bar{\tau}_i, \bar{\theta}_i) & = L_{loc}(\bar{\tau}_i, \bar{\theta}_i) + L_{obj}(\bar{\tau}_i, \bar{\theta}_i) \\
            & = \left[\left(\Delta \hat{\bar{\tau}}_i - \Delta \bar{\tau}_i\right)^2 + \left(\Delta \hat{\bar{\theta}}_i - \Delta \bar{\theta}_i\right)^2\right] \\
            & - \rho \left[C_i log\left(\hat{C}_i\right) + \left(1-C_i\right) log\left(1-\hat{C}_i\right)\right], 
    \end{split}
\end{equation}
where $\rho$ is a hyper-parameter to balance the loss of the objectness and localization for a better training performance. The ground truth of the confidence $C_i$ is set to 1 if there is a scatter in the scanning region of the anchor, otherwise it is set to 0.
In the meantime, the anchors can be classified into positive and negative anchors according to whether there is a scatter in their scanning region.
The positive anchors are trained with the full loss function in (\ref{training_target}).
For the negative anchors, they will only be trained with the objectness loss to avoid false detections.
Thus, for the whole channel matrix with multiple anchors, the total training loss function of the network $f(\cdot)$ will be expressed as the sum of the losses of all the anchors:
\begin{equation}
    \label{total_training_target}
    \begin{split}
        L & = \sum_{i \in \mathbb{P}} \left[L_{loc}(\bar{\tau}_i,\bar{\theta}_i) + L_{obj}(\bar{\tau}_i,\bar{\theta}_i) \right]+ \sum_{i \in \mathbb{N}} L_{obj}(\bar{\tau}_i,\bar{\theta}_i) \\
            & = \sum_{i \in \mathbb{P}} \left[\left(\Delta \hat{\bar{\tau}}_i - \Delta \bar{\tau}_i\right)^2 + \left(\Delta \hat{\bar{\theta}}_i - \Delta \bar{\theta}_i\right)^2 \right] \\
            & - \rho \sum_{i = 1}^{N_a} \left[C_i log\left(\hat{C}_i\right) + \left(1-C_i\right) log\left(1-\hat{C}_i\right)\right],
    \end{split}
\end{equation}
where $\mathbb{P}$ and $\mathbb{N}$ are the sets of positive and negative anchors, respectively.

\subsubsection{Adaptive ghost target removal}

As a common challenge in traditional methods like MUSIC-based \cite{chen2023music, lu2024isac4d} or ToF-based \cite{shi2022devicefree} methods, the ghost targets can be easily mis-detected as real scatters, which can significantly influence the localization performance.
This arises because the communication paths can generate overlap or interference in the channel matrix with the multi-path effect and estimation noise, producing ghost targets that conventional signal processing methods fail to distinguish.
In particular, without prior knowledge of the number of scatters, ghost targets can be easily misidentified, greatly impacting the localization accuracy.

In contrast, the proposed CSIYOLO framework can adaptively estimate the confidence of each detected scatter and remove ghost targets via a simple thresholding strategy.
Specifically, after the coordinate offsets and confidence of each scatter are obtained by the detection network $f(\cdot)$, it will first transform the relative offsets into absolute coordinates.
Then, the output candidate scatters are coarsely filtered using a low confidence threshold $t_1$ (e.g., 0.5), with detections below $t_1$ discarded.
Remaining scatters are subsequently merged based on spatial proximity, where nearby detections are grouped and replaced by a single scatter $\left(\hat{\bar{\tau}}^{'}, \hat{\bar{\theta}}^{'}\right)$ as follows:
\vspace{-4mm}
\begin{equation}
    \label{merge}
    \begin{aligned}
        & \left(\hat{\bar{\tau}}^{'}, \hat{\bar{\theta}}^{'}\right) 
        = \left\{ \left(\hat{\bar{\tau}}_j, \hat{\bar{\theta}}_j \right) : 
        j \in \mathbb{I}, \, 
        \hat{C}_j = \max_{i \in \mathbb{I}} \hat{C}_i \right\}, \\
        & \mathbb{I} = \Bigl\{ i : \forall i_1, i_2 \in \mathbb{I}, \,
        \left(\hat{\bar{\tau}}_{i_1} - \hat{\bar{\tau}}_{i_2}\right)^2
        + \left(\hat{\bar{\theta}}_{i_1} - \hat{\bar{\theta}}_{i_2}\right)^2 
        < t_d \Bigr\},
    \end{aligned}
\end{equation}
with $t_d$ denoting a threshold that controls the merging scope. Finally, a fine-grained threshold $t_2$, adaptively set to one-third of the average confidence of remaining candidates, is applied to further refine detections and eliminate ghost targets.
In fact, thanks to the strong detection capability of CSIYOLO, the confidences of possible scatters are highly concentrated, allowing ghost targets easily removed with simple thresholding.

\subsubsection{Training and inference procedure}
In summary, the proposed anchor-based scatter parameter detection method is composed of training and inference stages as outlined in \textbf{Algorithm \ref{alg_anchor}}.
During training, channel samples are fed into the anchor-based detection network $f(\cdot)$ to generate estimated scatter coordinates, and the loss function in (\ref{total_training_target}) is computed and backpropagated to update network weights.
During inference, the estimated channel matrix is processed by the trained network to produce candidate coordinates, which are then refined through the aforementioned target removal strategy to yield the final detected scatter coordinates.
After that, the real scatter parameters are recovered from these coordinates to support the following localization.

\begin{algorithm}[!t]
\caption{Anchor-based scatter parameter detection}
\label{alg_anchor}
\begin{algorithmic}[1]
    \REQUIRE
    \STATE \textbf{Initialize:} Prepare training datasets and parameters;
    \FOR{epoch = 1 to $N_{epoch}$}
        \FOR{batch = 1 to $N_{batch}$}
            \STATE \textbf{Forward:} Pass the angular-delay domain channel $\bar{\mathbf{H}}$ through the anchor-based detection network $f(\cdot)$ to obtain scatter coordinate offsets according to (\ref{forward});
            \STATE \textbf{Backpropagation:} Calculate loss $L$ with (\ref{total_training_target}) and backpropagate gradients to update network weights;
        \ENDFOR
    \ENDFOR
    \STATE \textbf{Output:} The trained anchor-based detection network $f(\cdot)$;
    \vspace{-4mm}
    \Statex \hrulefill
    \ENSURE
    \STATE \textbf{Initialize:} The estimated channel matrix $\bar{\mathbf{H}}$ in the angular-delay domain;
    \STATE \textbf{Forward:} Pass $\bar{\mathbf{H}}$ through the trained detection network $f(\cdot)$ to obtain candidate coordinate offsets via (\ref{forward});
    \STATE \textbf{Coordinate transformation:} Transform the relative coordinate offsets to the absolute scatter coordinates;
    \STATE \textbf{Coarse filtering:} Remove candidates with $\hat{C}_i < t_1$;
    \STATE \textbf{Merge:} Merge candidates with nearby coordinates according to (\ref{merge});
    \STATE \textbf{Fine filtering:} Remove candidates with $\hat{C}_i < t_2$, where $t_2$ is set to one-third of the average confidence of remaining candidates;
    \STATE \textbf{Return:} The estimated scatter coordinates $\left\{\left(\hat{\bar{\tau}}_l, \hat{\bar{\theta}}_l\right)\right\}_{1 \leq l \leq \hat{N}_s}$, where $\hat{N}_s$ is the number of remaining candidates.
\STATE \textbf{Output:} The estimated scatter parameters $\left\{\left(\hat{\tau}_l, \hat{\theta}_l\right)\right\}_{1 \leq l \leq \hat{N}_s}$ recovered according to (\ref{parameter_from_coordinate}).
\end{algorithmic}
\end{algorithm}

\subsection{Stage 2: CSI-based scatter localization with a single user}

\begin{figure}[!t]
\centering
\vspace{-2mm}
\includegraphics[width=0.9\linewidth]{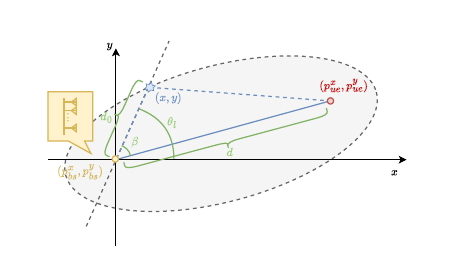}
\vspace{-3mm}
\caption{The diagram of the ellipse model of the scatter localization method.}
\label{fig_ellipse} 
\vspace{-5mm}
\end{figure}

With the estimated scatter parameters from the anchor-based detection method, the position of each scatter can be estimated based on the ellipse model as follows:
\begin{itemize}
    \item Firstly, the transmission delay $\tau_l$ represents the difference in distance between the reflected path and the direct path. Since the length of the direct path is determined by the positions of the BS and the UE and remains fixed for a given scatter, a specific transmission delay $\tau_l$ corresponds to a set of possible scatter positions $\mathbf{p}_{s,l} \triangleq (x, y)$ that form an ellipse with the BS and the UE as its foci. This relationship can be expressed as:
    \begin{equation}
    \label{ellipse}
    \begin{aligned}
        \sqrt{\left(x - p_{bs}^x\right)^2 + \left(y - p_{bs}^y\right)^2} & + \sqrt{\left(x - p_{ue}^x\right)^2 + \left(y - p_{ue}^y\right)^2} \\
        & = c\tau_l + \Vert \mathbf{p}_{bs} - \mathbf{p}_{ue}\Vert.
    \end{aligned}
    \end{equation}
    \item Secondly, the AoD of the $l$-th path $\theta_l$ can be used to further constrain the possible positions of the scatter. Specifically, $\theta_l$ represents the relative direction of the scatter with respect to the vertical axis of the antenna array at the BS. As illustrated in Fig.~\ref{fig_ellipse}, the line originating from the BS at angle $\theta_l$ intersects the ellipse at two points, one of which lies outside the antenna sector and can be discarded. The remaining intersection uniquely determines the position of the scatter, which can be formulated as:
    \vspace{-2mm}
    \begin{equation}
    \label{intersect}
        \frac{y}{x} = \tan(\theta_l)
    \end{equation}
\end{itemize}

With (\ref{ellipse}) and (\ref{intersect}), the position of the $l$-th scatter can be finally derived as:
\begin{equation}
    \label{position}
    \begin{split}
        x & = d_0 \cos(\theta_l) + p^x_{bs}, \\
        y & = d_0 \sin(\theta_l) + p^y_{bs}, \\
    \end{split}
\end{equation}
where $d_0$ follows the following equations:
\begin{equation}
    \label{d_0}
    \begin{split}    
        d_0 & = \frac{c \tau_l \left(2 d + c \tau_l\right)}{2 \left(c \tau_l + d - d \cos(\beta)\right)} \\
        d & = \Vert \mathbf{p}_{bs} - \mathbf{p}_{ue}\Vert, \\
        \cos(\beta) & = \cos(\theta_l) \frac{p_{ue}^x - p_{bs}^x}{d} + \sin(\theta_l) \frac{p_{ue}^y - p_{bs}^y}{d}.
    \end{split}
\end{equation}
Thus, through applying the aforementioned derivations to the estimated scatter parameters $\left\{ \left( \hat{\tau}_l, \hat{\theta}_l \right) \right\}_{1 \leq l \leq \hat{N}_s}$, the positions of the scatters can be finally obtained.

In fact, from a broader perspective, the channel matrix in the angular-delay domain can be regarded as a latent representation of the physical 2D coordinate system corresponding to real-world scatter locations.
The preceding derivations establish a mapping function that links the coordinates on the channel matrix to the actual spatial positions of the scatters. This insight highlights the validity and potential of the proposed CSI-based scatter sensing framework.

\section{Network design and training strategy}
In this section, we detail the design of the proposed anchor-based detection network $f(\cdot)$ and the training strategy within the CSIYOLO framework. 
As highlighted in the previous discussion, task-oriented optimizations are necessary and introduced to better tailor the YOLO-inspired architecture to the characteristics of the channel matrix, enhancing localization accuracy in the scatter sensing task.

\vspace{-2mm}
\subsection{Network architecture}
\begin{figure*}
\centering
\includegraphics[width=0.97\linewidth]{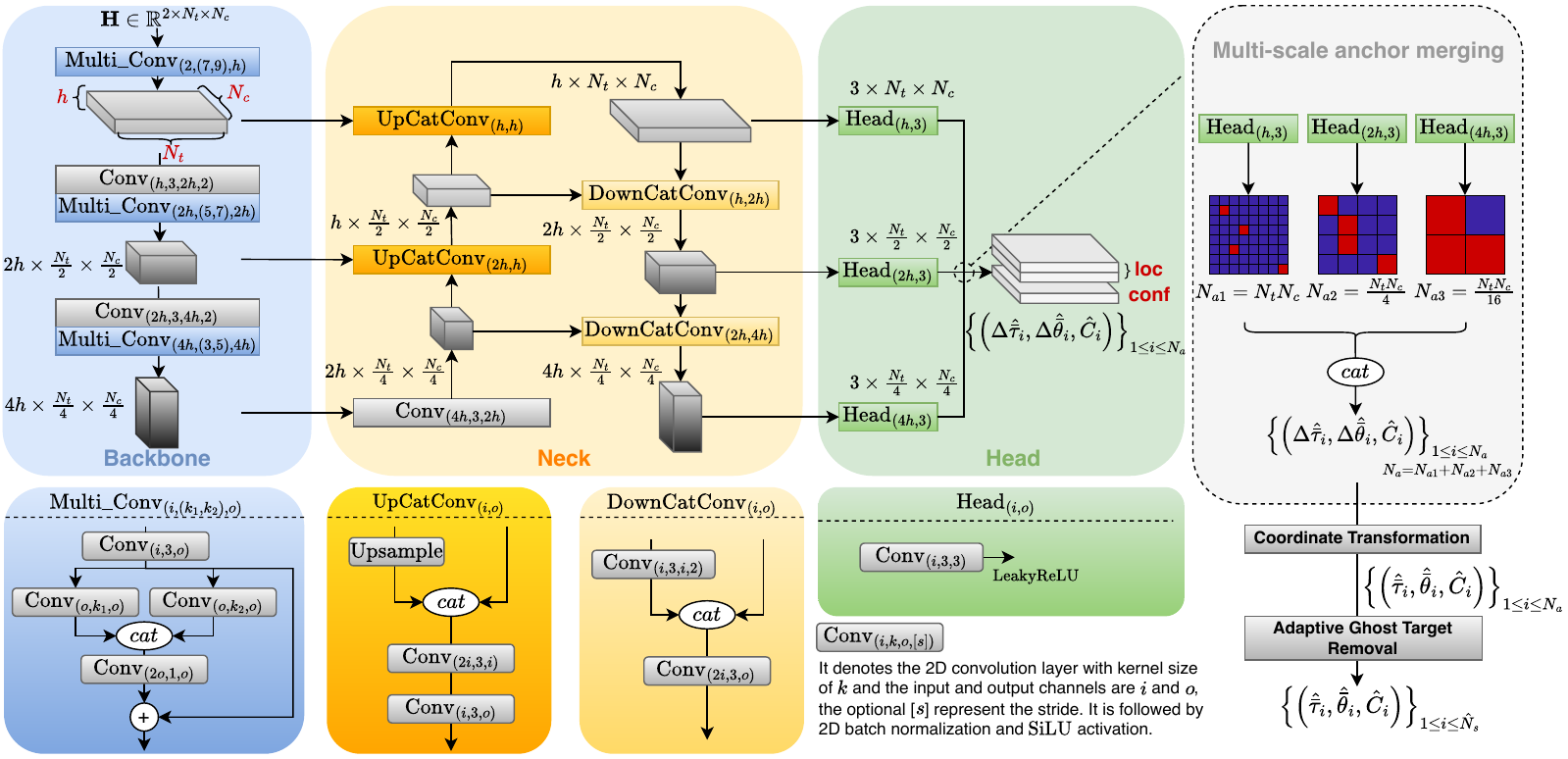}
\vspace{-2mm}
\caption{The diagram of the extendable network structure together with the multi-scale anchor detection merging. \label{network}}
\vspace{-4mm}
\end{figure*}

To achieve efficient feature extraction, we propose an extendable network architecture built upon the U-Net framework \cite{ronneberger2015unet}.
The U-Net design is particularly well-suited for precise localization tasks, as its skip connections preserve rich spatial information.
By integrating the multi-resolution mechanism \cite{lu2020multi} into the U-Net, the proposed network can effectively capture scatter features across multiple scales.
Furthermore, circular convolutions and convolution factorization are incorporated to enhance localization accuracy while reducing computational complexity under resource constraints, thus enabling a flexible and efficient network design.
As illustrated in Fig. \ref{network}, the proposed architecture is composed of three main components: the backbone, neck, and head.

\begin{itemize}
    \item \textbf{Backbone}. 
    The backbone serves as the feature extraction module, composed of convolutional layers. Successive convolutions increase the number of feature channels and expand the feature space. To handle varying scales of channel features in the angular-delay domain, multi-scale convolution layers \cite{lu2020multi} with different kernel sizes are employed, and their outputs are merged via concatenation. In this way, the channel matrix is transformed into a high-dimensional feature space that more effectively captures and represents the channel characteristics. Moreover, the depth $h$ of the hidden feature space can be adaptively adjusted according to computational resources, providing an extendable architecture that flexibly balances localization accuracy and efficiency.
    \item \textbf{Neck}.
    Following the backbone, the neck module integrates features across different dimensions, thereby constructing a more holistic representation of the channel. By employing both downsampling and upsampling operations, features at multiple dimensions and resolutions are effectively fused into a unified feature map. This enriched representation facilitates the identification of communication paths of varying sizes.
    \item \textbf{Head}. 
    Finally, the head module projects the fused channel features into the output parameter space. Specifically, multiple heads of varying sizes are attached to the multi-scale feature maps generated by the neck, representing multi-scale anchor patterns with different counts and scanning resolutions, as shown in Fig. \ref{network}.
    For an output shape of $(3 \times a \times b)$, the number of anchors is $a \times b$. 
    Each anchor covers a scanning region of size $\frac{2 N_t}{a} \times \frac{2 N_c}{b}$ and outputs a three-element vector containing estimated scatter coordinates and confidence.
    With different output sizes, scatters can be detected at multiple resolutions.
    By merging outputs across different scales, scatter parameters can be estimated with enhanced accuracy.

\end{itemize}
Through the aforementioned network design, the channel matrix is analyzed at multiple scales, allowing the features to be extracted with greater accuracy.

\subsection{Circular convolution}
To further enhance localization accuracy, we utilize the \textit{circular convolution} in the detection network.
Unlike the commonly used zero-padding in convolution layers, circular padding is adopted along the angular dimension of the 2D channel matrix. 
This choice is motivated by the fact that the angular dimension is inherently periodic \cite{liu2024circular}, whereas zero-padding disrupts this periodicity, leading to information loss and degraded localization performance.

\begin{figure}[!t]
\includegraphics[width=\linewidth]{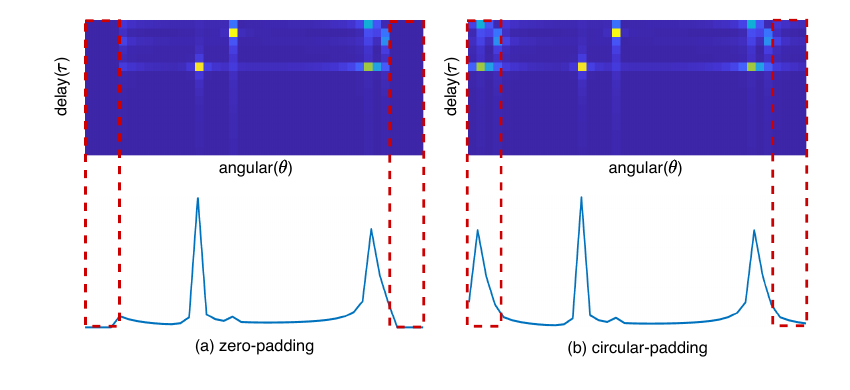}
\vspace{-4mm}
\caption{The diagram of different padding strategies. The red dashed box marks the padded region. Compared with zero-padding, circular padding better preserves boundary information.}
\label{fig_circular} 
\vspace{-3mm}
\end{figure}

As illustrated in Fig. \ref{fig_circular}, with conventional zero-padding, the circular nature of the channel matrix along the angular dimension is broken, and scatters located near the boundary are prone to being overlooked due to the abrupt cutoff at the edges. 
In contrast, circular padding preserves the periodicity of the channel matrix, enabling more energy to be captured from boundary regions during convolution. 
This can increase the likelihood of detecting boundary scatters and improve their localization accuracy.
For the delay dimension, zero-padding is still employed to maintain the size of the channel matrix.

\subsection{Convolution factorization}
\begin{figure}[!t]
\centering
\vspace{-2mm}
\includegraphics[width=0.95\linewidth]{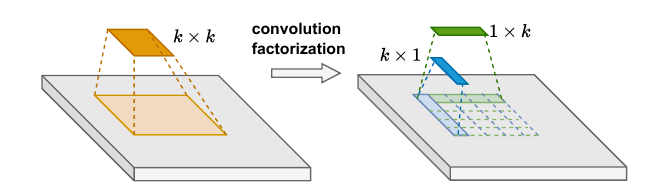}
\caption{The diagram of convolution factorization. }
\label{fig_factorization} 
\vspace{-3mm}
\end{figure}

Furthermore, to reduce network complexity while maintaining localization accuracy, we introduce the \textit{convolution factorization} technique \cite{szegedy2015Inception} in the detection network. 
As shown in Fig. \ref{fig_factorization}, a standard 2D convolution can be decomposed into two successive one-dimensional (1D) convolutions with orthogonal kernel orientations. 
This decomposition achieves a comparable receptive field while significantly reducing the number of parameters and computational costs.

For example, a $k \times k$ 2D convolution can be factorized into a $1 \times k$ convolution followed by a $k \times 1$ convolution. 
Through this way, the computational complexity decreases from $O(k^2)$ to $O(2k)$, thereby reducing resource requirements while preserving the spatial structure of the feature maps. 
Besides, the factorization enables the insertion of non-linear activation functions between the two 1D convolutions, which further enhances the non-linearity and expressive power of the network, ultimately improving localization performance.

Overall, convolution factorization provides a favorable trade-off between computational efficiency and localization accuracy, making the network more suitable and flexible for deployment in resource-constrained environments.

\vspace{-2mm}
\subsection{Noise injection training strategy}
To further enhance the robustness of the detection network against channel estimation errors, we propose a \textit{noise injection} training strategy. Instead of relying on estimated channels for training, which may introduce considerable data preparation overhead, we directly inject Gaussian noise into the channel matrix during training.

In this strategy, the noise level is progressively increased as training advances. 
Such a gradual scheme prevents the destruction of channel characteristic learning at the early stage while enabling the network to adapt to higher noise levels in later stages. 
With the maximum noise level set to $\sigma_{max}$, the noisy channel matrix generated at each step covers a relatively wide range of noise intensities from $0$ to $\sigma_{max}$, further improving the robustness of the network against diverse channel estimation errors. 
Formally, the injected noise is modeled as:
\begin{equation}
    n(t) \sim \mathcal{N}(0, \sigma^2(t)),
\end{equation}
where the variance $\sigma^2(t)$ is randomly sampled from a uniform distribution within the range of $[0, \sigma_{max}^2(t)]$. Here, the maximum noise variance $\sigma_{max}^2(t)$ is designed to increase with the training step $t$:
\begin{equation}
    \sigma_{max}^2(t) = \sigma_0^2 + \gamma t,
\end{equation}
where $\sigma_0^2$ is the initial noise variance, and $\gamma$ is a hyper-parameter that controls the rate of increase in noise variance.

This formulation simulates the growing uncertainty inherent in channel estimation. 
By incorporating such noise injection into training, the network learns to handle imperfect channel conditions and becomes more resilient to estimation errors. 
Simulation results demonstrate that this strategy effectively enhances localization performance under noisy channels.

\section{Simulation results and analysis}

In this section, we present the simulation settings and experimental results to validate the effectiveness of the proposed framework. 

\subsection{Experimental settings}

\begin{table}
    \footnotesize
    \centering
    \caption{Dataset generation parameters.}
    \label{tab_dataset}
    \begin{tabular}{c|c}
        \hline
        \textbf{Parameter} & \textbf{Value} \\
        \hline
        Downlink center frequency $f_0$& 28 GHz \\
        Frequency spacing $\Delta f$& 100 MHz \\
        Number of sub-carriers $N_c$ & 1024 \\
        Number of transmit antennas $N_t$ & 64 \\
        Number of scatters $N_s$ & $\left[5, 10\right]$ \\
        Area size & 100m$\times$100m\\
        Number of left rows $N_c^{'}$& 64 \\
        \hline
    \end{tabular}
\vspace{-5mm}
\end{table}
The parameters used for dataset generation are summarized in Table \ref{tab_dataset}. We consider a single-cell scenario as shown in Fig. \ref{fig_isac_system} with the BS located at $(0,0)$. The UE and scatters are randomly distributed within a rectangular area centered at $(50\text{m}, 0)$, with a size of $(100\text{m}, 100\text{m})$. The BS is equipped with $N_t = 64$ antennas arranged in a ULA configuration, while the UE is equipped with a single antenna. The number of scatters is randomly chosen as $N_s \in [5,10]$, forming the communication paths between the BS and the UE. 

Besides, the channel matrix in the angular-delay domain is first truncated along the delay dimension to remove near-zero entries, retaining only the first $N_c^{'} = 64$ rows for subsequent processing. This truncation is justified by the fact that the transmission delay of communication paths is inherently limited, and such preprocessing has been widely adopted in CSI-related studies \cite{wen2018deep, zhang2024continuous, liu2024circular}. In addition, the first row of the channel, which corresponds to the LOS path, is removed to mitigate the dominance of its large energy. These operations can effectively narrow the network's region of interest and further reduce computational complexity.

The dataset is randomly divided into training, validation, and testing sets with 10,000, 3,000, and 2,000 samples, respectively. Training is performed using the SGD optimizer with an initial learning rate of $1 \times 10^{-2}$. The batch size is set to 64, and the training process runs for 100 epochs. 
The balance coefficient $\rho$ in the loss function (\ref{total_training_target}) is set to 1.
All experiments are conducted on an NVIDIA RTX P100 GPU.

To evaluate localization performance, we define a scatter to be correctly localized if the Euclidean distance between its estimated position and the ground-truth position is smaller than 1 m. 
Based on this criterion, the probability of detection $P_d$ (equivalently, recall $R$) and the F1-score $F_1$ are adopted as evaluation metrics:
\vspace{-1mm}
\begin{equation}
    \label{accuracy}
    P_d(r) = R(r) = \frac{1}{N_s} \sum_{l=1}^{N_s} \mathbbm{1}\left( \left[\min_{1 \leq l^{'} \leq \hat{N}_s}\Vert \hat{\mathbf{p}}_{s,l^{'}} - \mathbf{p}_{s,l} \Vert \right]< r\right),
\end{equation}
\vspace{-2mm}
\begin{equation}
    \label{recall}
    P(r) = \frac{1}{\hat{N}_s} \sum_{l^{'}=1}^{\hat{N}_s} \mathbbm{1}\left( \left[\min_{1 \leq l \leq N_s}\Vert {\mathbf{p}}_{s,l} - \hat{\mathbf{p}}_{s,l^{'} } \Vert \right]< r\right),
\end{equation}
\vspace{-2mm}
\begin{equation}
    \label{f1}
    F_1(r) = 2 \cdot \frac{P(r) \cdot R(r)}{P(r) + R(r)},
\end{equation}
where $r$ is the detection threshold and is set to 1 in this experiment.
The probability of detection $P_d$, which is equivalent to recall $R$, measures the accuracy of detection, with higher values indicating that a larger proportion of true scatters are correctly localized. 
The F1-score $F_1$, defined as the harmonic mean of precision and recall, provides a balanced evaluation between missed detections and false alarms.
A higher F1-score $F_1$ indicates better overall performance, corresponding to fewer missed scatters and ghost targets.
Therefore, these two metrics can reflect the localization performance.

Furthermore, we also evaluate the localization accuracy for detected scatters with the root mean square error (RMSE):
\vspace{1mm}
\begin{equation}
    \label{rmse}
    \text{RMSE} = \sqrt{\frac{1}{| \mathbb{D} |} \sum_{\mathbf{p}_{s,l} \in \mathbb{D}} \min_{1 \leq l^{'} \leq \hat{N}_s} \Vert \hat{\mathbf{p}}_{s,l^{'}} - \mathbf{p}_{s,l}\Vert^2},
\end{equation}
where $\mathbb{D}$ is the set of detected scatters. It can further help to analyze the localization performances.

For comparison, we implement two widely adopted baseline methods: the MUSIC and DFT-based approach \cite{kim2019MUSIC, chen2023music, lu2024isac4d} and the LTD method \cite{mou2023mmwave}:
\begin{itemize}
    \item \textbf{MUSIC-FFT} \cite{lu2024isac4d}: As commonly used in radar signal processing, the MUSIC algorithm is employed in the angular domain to estimate AoDs, while a DFT-based method is used in the delay domain to estimate path delays. Estimated parameters are then combined to derive scatter positions.
    \item \textbf{LTD} \cite{mou2023mmwave}: The low-rank tensor decomposition method jointly exploits angular-delay domain information by leveraging spatial and temporal correlations. Scatter parameters are estimated by detecting peaks over 2D grids.
\end{itemize}
However, both methods typically require prior knowledge of the number of scatters in the environment to ensure reliable performance. 
In contrast, our proposed framework adaptively estimates the confidence of each detection and effectively eliminates ghost targets. 
Furthermore, we compare CSIYOLO with the direct application of generic YOLO structures as in existing studies \cite{li2020yolo,han2020yolo}, to highlight the effectiveness of our optimized network design.

\subsection{Localization performance}
In this subsection, we present the localization performance of our proposed CSIYOLO framework.
The convolution factorization and circular convolution are employed by default.
It should be noted that the number of scatters is not known in advance for our proposed method and they are estimated adaptively during detection.
For the MUSIC-FFT, the number of scatters is assumed to be known in advance.

\begin{figure}
\centering
\includegraphics[width=0.9\linewidth]{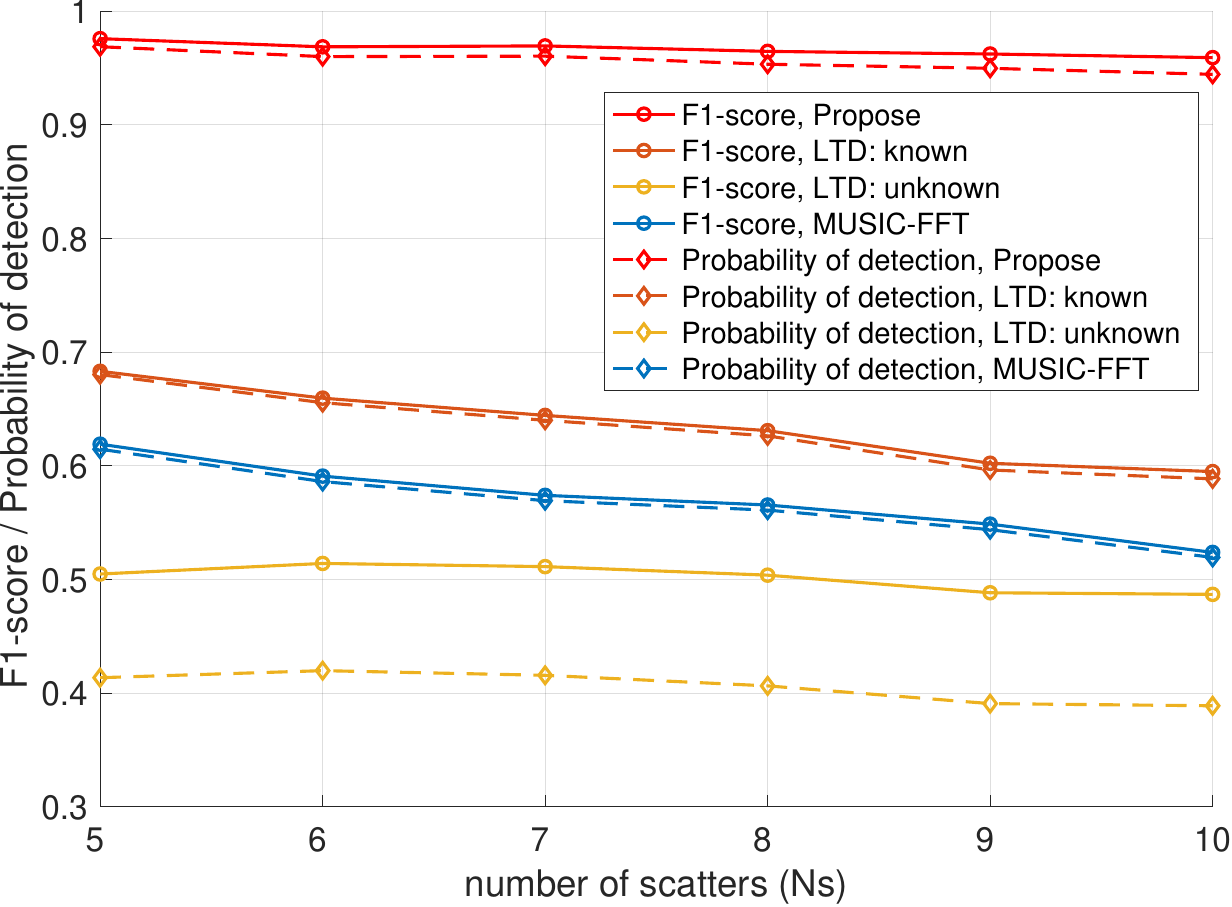}
\vspace{-2mm}
\caption{Localization performance under different numbers of scatters. $h$ is set to 5 for the proposed method.}
\label{fig_fixed}
\vspace{-3mm}
\end{figure}

As shown in Fig. \ref{fig_fixed}, our proposed CSIYOLO method can achieve the best localization performance with the highest probability of detection and F1-score.
It can also be observed that, for each of the proposed method, MUSIC-FFT, and LTD with known numbers of scatters, the F1-score is close to its corresponding probability of detection, indicating that the estimated number of scatters is generally consistent with the ground truth.
Considering that the number of scatters is not known in advance for the proposed method, this result further highlights its superiority.

\begin{figure}
\includegraphics[width=0.95\linewidth]{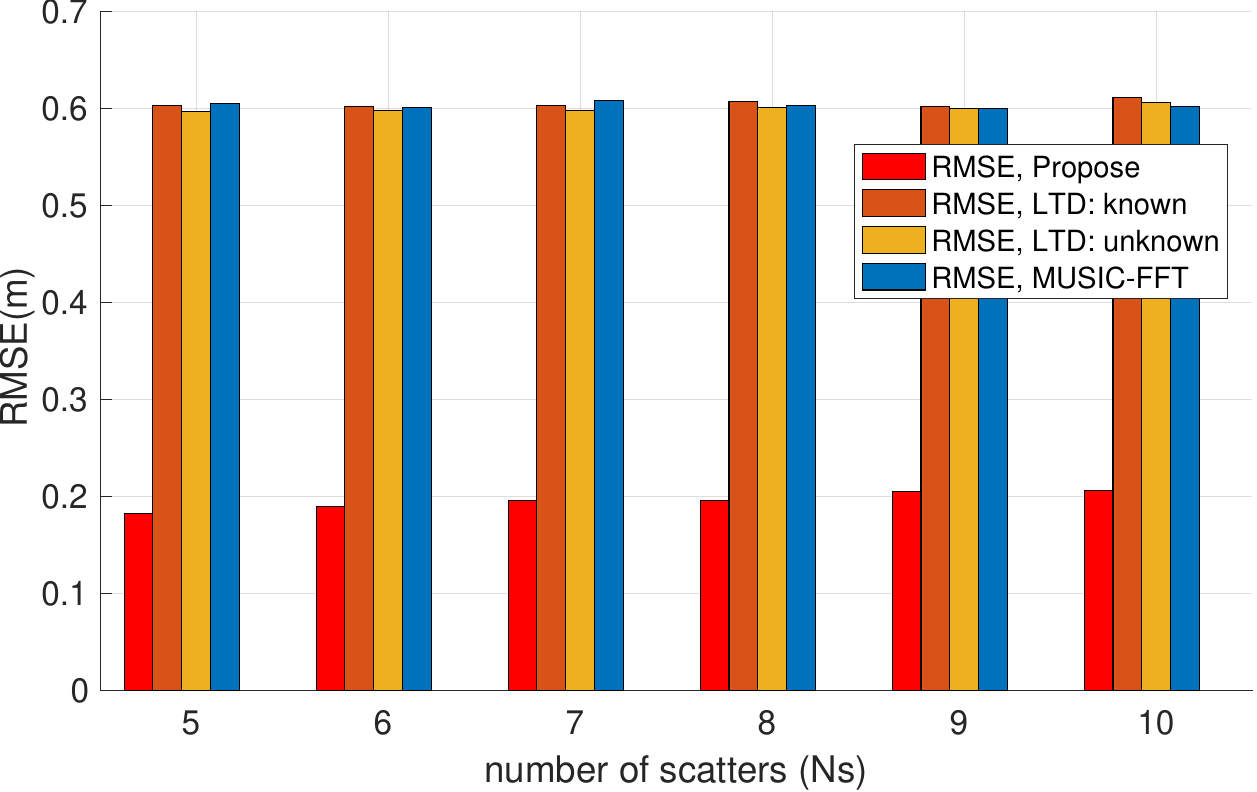}
\vspace{-2mm}
\caption{Localization RMSE performance under different numbers of scatters. $h$ is set to 5 for the proposed method. \label{fig_rmse}}
\vspace{-3mm}
\end{figure}

The localization RMSE is presented in Fig. \ref{fig_rmse}, showing that the proposed method achieves the lowest RMSE, even below 0.2 m, outperforming the baseline methods by a significant margin.
The baseline methods exhibit similar RMSE values, as they rely on grid-based detection, which is inherently limited by the grid resolution and cannot achieve high localization accuracy.
In contrast, our proposed method can predict continuous localization results, overcoming the limitations of the grid resolution and further improving localization accuracy.

\begin{figure}
\includegraphics[width=\linewidth]{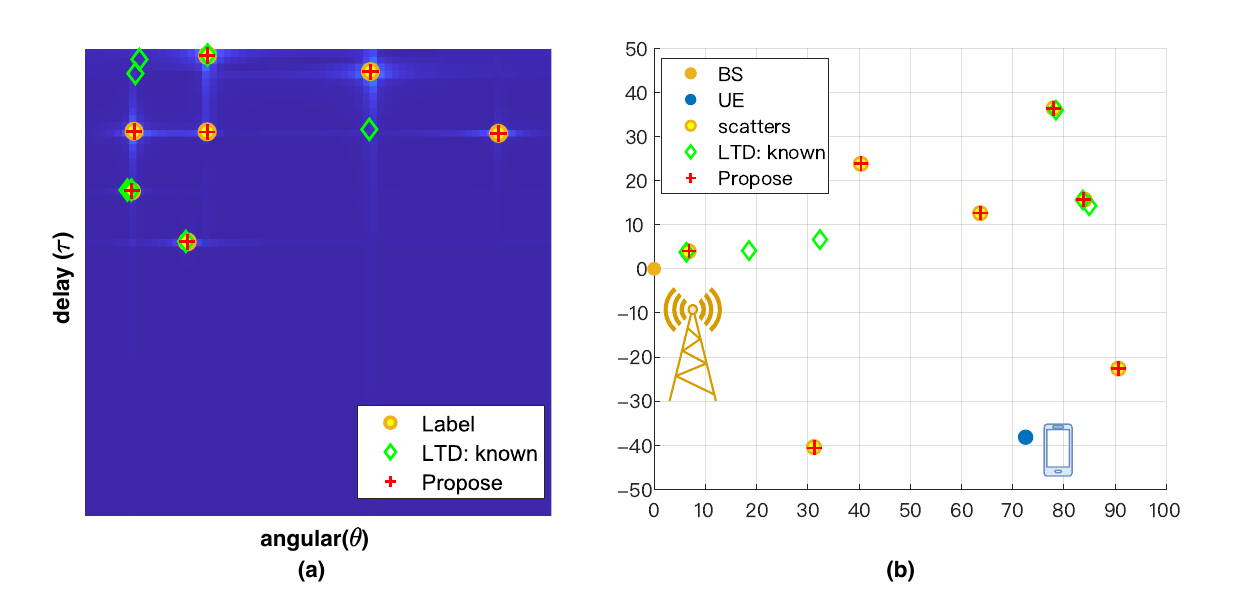}
\vspace{-4mm}
\caption{Localization examples of different methods. (a) Detection performance of scatter parameters on the angular-delay domain channel. (b) Localization performance of scatters on the 2D coordinate space.}
\label{fig_example}
\vspace{-3mm}
\end{figure}

Fig. \ref{fig_example} illustrates an example of the localization results. It can be observed that the LTD method is prone to confusion by ghost targets caused by multi-path interference. By comparison, the proposed method demonstrates superior parameter detection and scatter localization accuracy.

\begin{figure}
\includegraphics[width=\linewidth]{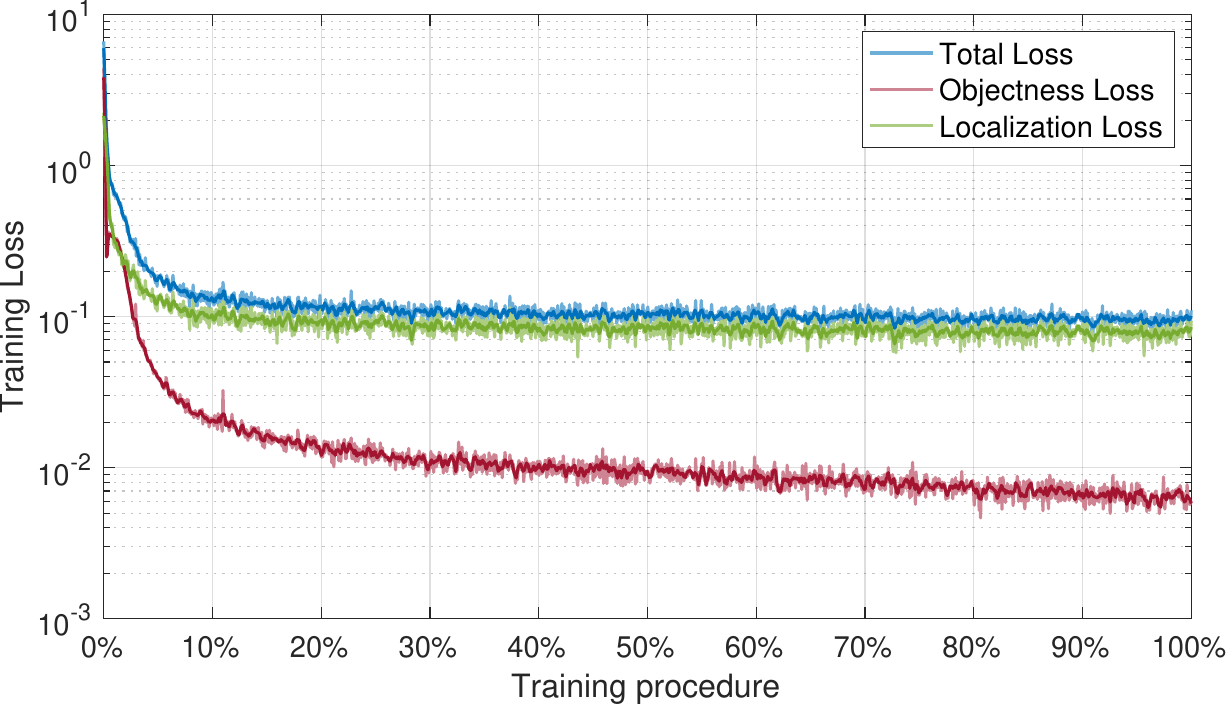}
\caption{Convergence performance of the proposed method through training.}
\label{fig_convergence_1}
\vspace{-3mm}
\end{figure}

\begin{figure}
\includegraphics[width=\linewidth]{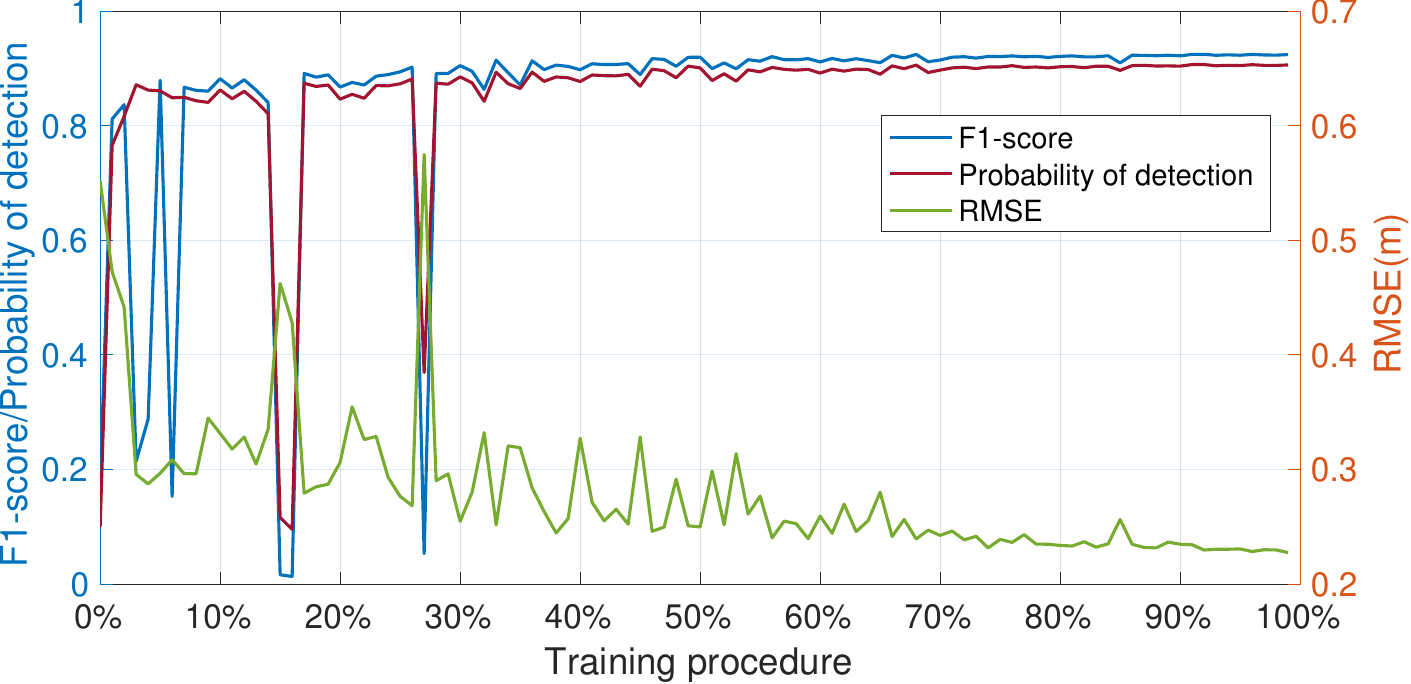}
\caption{Validation performance of the proposed method through training.}
\label{fig_convergence_2}
\vspace{-3mm}
\end{figure}

Furthermore, the convergence performance of the proposed method is presented in Fig. \ref{fig_convergence_1} and \ref{fig_convergence_2}. 
It can be seen that both the objectness and localization training loss of the proposed method can achieve stable convergence through training.
Meanwhile, after fluctuations in the early stage of training, the validation performance also shows a steady improvement.

\subsection{Effectiveness of network design}

\begin{figure*}
\centering
\includegraphics[width=0.97\linewidth]{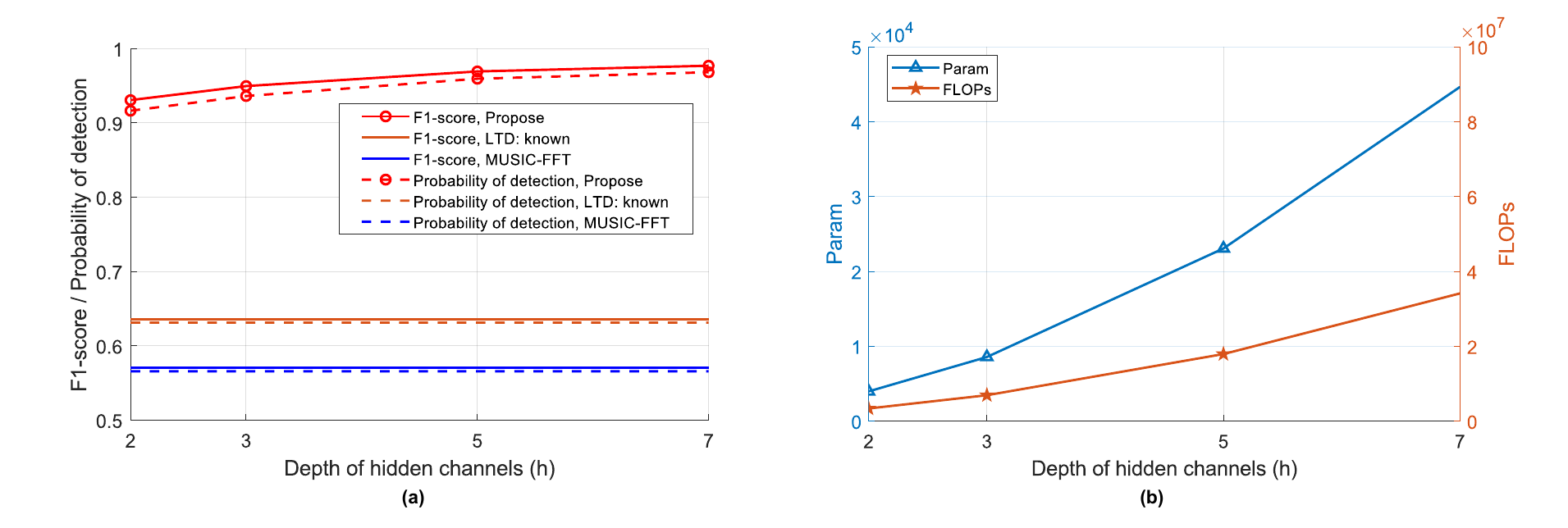}
\vspace{-3mm}
\caption{The performance and complexity comparison of the proposed methods with different depths of hidden channels. (a) Performance comparison of different depths of hidden channels. (b) Complexity comparison of different depths of hidden channels.}
\label{fig_complexity}
\vspace{-3mm}
\end{figure*}

In this subsection, we conduct ablation studies to validate the effectiveness of the proposed anchor-based detection network design, including the extendable hidden channel structure, circular convolution, and convolution factorization. 
First, Fig. \ref{fig_complexity}(a) shows the localization performance of the proposed method under different hidden channel depths $h$, while Fig. \ref{fig_complexity}(b) compares the corresponding model complexity in terms of the number of parameters (Param) and floating-point operations (FLOPs). 
Results show that as the hidden channel depth decreases, the localization performance degrades only marginally while still significantly outperforming the baseline methods. 
Meanwhile, the computational complexity is substantially reduced. 
This demonstrates that the proposed method can flexibly extend the hidden channel depth according to device resource constraints, achieving an effective trade-off between performance and complexity.

\begin{figure}
\centering
\includegraphics[width=0.95\linewidth]{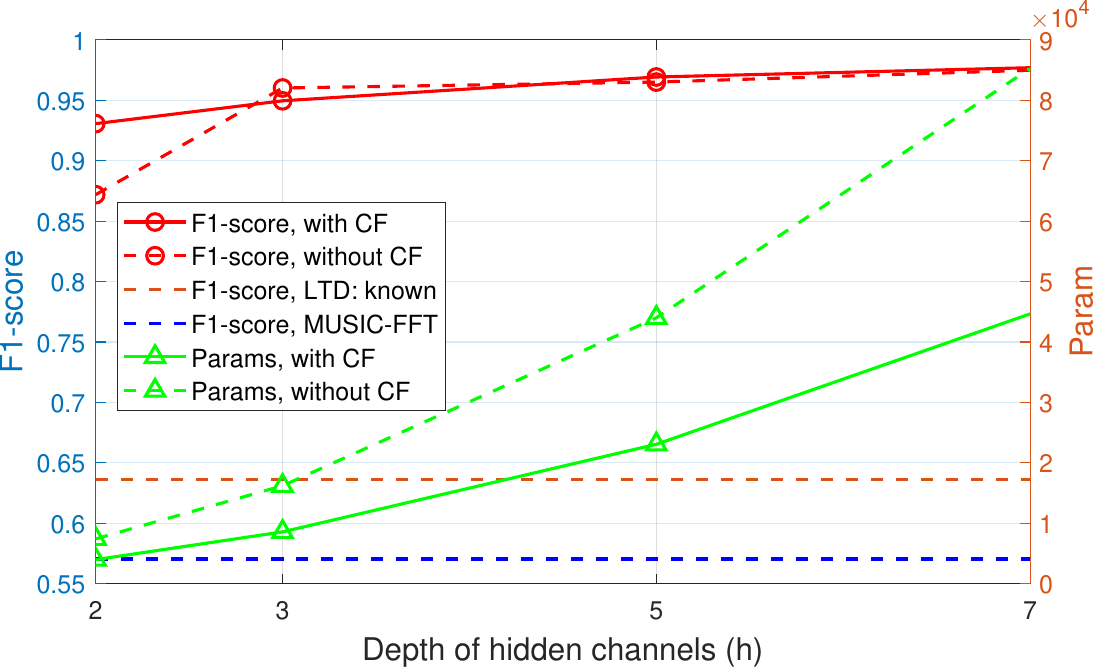}
\caption{Comparison of localization F1-score and complexity under random numbers of scatters. ``CF'' represents the convolution factorization. \label{fig_complexity_factorization}}
\vspace{-3mm}
\end{figure}

Furthermore, we evaluate the effectiveness of the convolution factorization method.
Fig. \ref{fig_complexity_factorization} compares the localization F1-score and network parameter amount for the proposed method with and without convolution factorization.
It can be seen that convolution factorization reduces the number of parameters by approximately 50\% while maintaining, or in some cases even improving, localization performance.
This means that with the additional non-linearity introduced by the factorization, the network can learn more sufficiently and achieve better localization performances.

The effectiveness of the circular convolution design is also confirmed. 
As shown in Table \ref{tab_circular}, circular convolutions preserve the periodicity of the channel matrix and consistently outperform conventional zero-padding convolutions in terms of localization performance in most cases.

\begin{table}
\centering
\footnotesize
\caption{Localization performance with and without circular convolutions under random numbers of scatter. \label{tab_circular}}
\begin{tabular}{c|c|c c c}
\hline
\multirow{2}{*}{\textbf{Width}} & \multirow{2}{*}{\textbf{Convolution}} & \multirow{2}{*}{\textbf{F1-score}} & \textbf{Probability} & \multirow{2}{*}{\textbf{RMSE(m)}} \\
 & & & \textbf{of detection} & \\
\hline
\multirow{2}{*}{2} & zero & 0.8839 & 0.8681 & 0.3518 \\
 & circular & \textbf{0.9307} & \textbf{0.9164} & \textbf{0.3056} \\
\hline
\multirow{2}{*}{3} & zero & 0.9322 & 0.9175 & 0.2371 \\
 & circular & \textbf{0.9496} & \textbf{0.9362} & \textbf{0.2317} \\
\hline
\multirow{2}{*}{5} & zero & 0.9642  & 0.8681 & \textbf{0.1941} \\
 & circular & \textbf{0.9693} & \textbf{0.9595} & 0.1949 \\
\hline
\multirow{2}{*}{7} & zero & 0.9615 &   0.9591 &  0.1685 \\
 & circular & \textbf{0.9770} & \textbf{0.9682} & \textbf{0.1579} \\
\hline
\end{tabular}
\vspace{-4mm}
\end{table}

To directly validate the effectiveness of the proposed network optimizations, we compare our model with the generic YOLOv5 structure \cite{yolov5github}, a widely adopted object detection framework. 
As shown in Table \ref{tab_yolov5}, the proposed method achieves substantially improved localization performance while reducing computational complexity with more than two orders of magnitude fewer parameters, demonstrating both the efficiency and effectiveness of the optimized design.

\begin{table}[!htb]
\footnotesize
\centering
\caption{Localization performance compared with YOLOv5.\label{tab_yolov5}}
\begin{tabular}{c|c|c c}
\hline
\textbf{Metric} & \textbf{Yolov5s} \cite{yolov5github} & \textbf{CSIYOLO}, $h$=5 & \textbf{CSIYOLO}, $h$=2 \\
\hline
\textbf{FLOPs} & 79.815M & 17.930M & \textbf{3.390M} \\
\textbf{Param} & ~7.028M & 23.069K & \textbf{3.977K} \\
\textbf{F1-score} & 0.8044 & \textbf{0.9693}  & 0.9307 \\
\textbf{${P_d}^*$} & 0.8504 & \textbf{0.9595} & 0.9164 \\
\textbf{RMSE(m)} & 0.3259 & \textbf{0.1949} & 0.3056 \\
\hline
\multicolumn{4}{l}{\footnotesize $^*$ $P_d$ represents the probability of detection.} 
\end{tabular}
\vspace{-3mm}
\end{table}

\subsection{Robustness on channel estimation}

\begin{figure}
\centering
\includegraphics[width=0.95\linewidth]{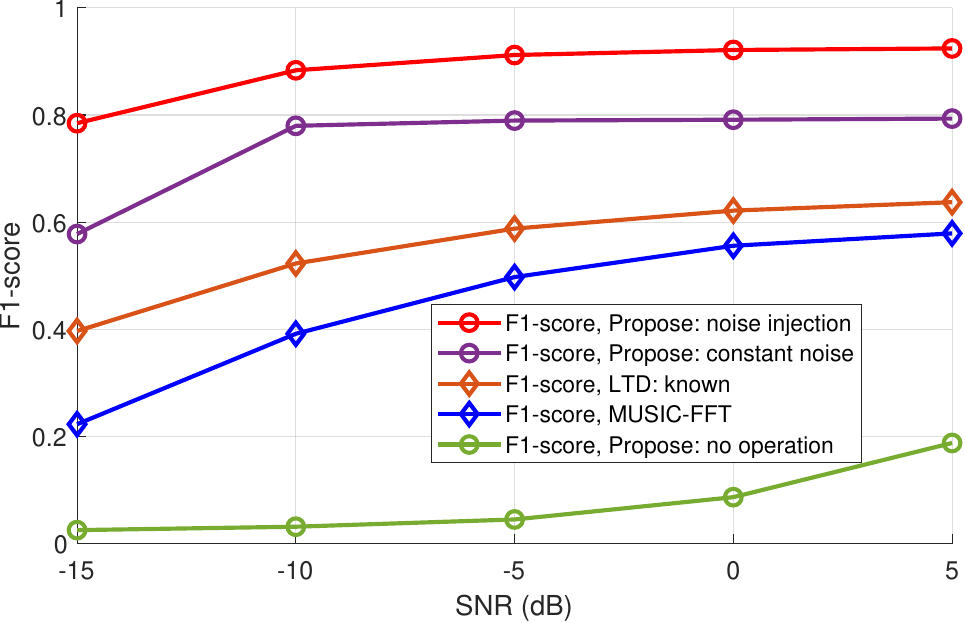}
\caption{Robustness on channel estimation errors of different methods under random numbers of scatters. $h$ is set to 5 for the proposed method.\label{fig_noise}}
\vspace{-3mm}
\end{figure}

Finally, we assess the effectiveness of the proposed noise injection training strategy by evaluating localization performance under different noise power levels, as illustrated in Fig. \ref{fig_noise}.

The results show that as the noise power increases, the localization performance of all methods deteriorates. 
Without noise-injection training, the proposed method suffers from severe performance degradation due to its inability to account for estimation errors. 
In contrast, with the noise injection strategy, the proposed method achieves substantially improved and stable localization performance across varying noise levels, demonstrating its robustness in practical scenarios.
Furthermore, the proposed linear noise injection scheme achieves even better performance compared with injecting constant noise, validating the robustness of the proposed training strategy.

\section{Conclusion}

This paper focused on the scatter sensing task in ISAC systems and proposed a novel CSIYOLO framework to localize environmental scatters from communication channels. 
By formulating scatter localization as an object detection problem, we developed an anchor-based detection method and an extendable network structure with specific optimizations, including circular convolution and convolution factorization, to improve both accuracy and computational efficiency. 
A noise injection training strategy was further introduced to enhance robustness against channel estimation errors. 
Simulation results demonstrated that CSIYOLO significantly outperforms baseline methods, while adaptively estimating scatter numbers and suppressing ghost targets for practical deployment. 
Future work will extend CSIYOLO to multi-user scenarios with information fusion to further improve localization performance.

\bibliographystyle{ieeetr}
\bibliography{main}

\end{document}